\documentclass[12pt]{article}
\usepackage{graphicx}

\newcommand{\ee}       {\mbox{$e^+ e^-$}}
\def\ra{\rightarrow}

\begin{document}
\bibliographystyle{hunsrt} 

\pagestyle{empty}

\renewcommand{\thefootnote}{\fnsymbol{footnote}}


\begin{flushright}
{\small
SLAC--PUB--10283\\
December 18, 2003\\}
\end{flushright}

\vspace{.8cm}

\begin{center}
{\large\bf
Higgs Coupling Measurements at a 1 TeV Linear Collider\footnote{Work supported by
Department of Energy contract  DE--AC03--76SF00515.}}

\vspace{1cm}

Timothy L. Barklow\\
Stanford Linear Accelerator Center, Stanford University\\
2575 Sand Hill Road, Menlo Park, CA  94025 USA\\

\end{center}

\vfill

\begin{center}
{\large\bf
Abstract }
\end{center}

\begin{quote}
Methods for extracting 
Higgs boson signals at  a 1~TeV center-of-mass energy $\ee$ linear collider are described.
In addition, estimates are given for the accuracy with which branching fractions can be measured 
for Higgs boson decays to 
$b\bar b$, $WW$, $gg$,  and $\gamma \gamma $.  
\end{quote}

\vfill

\begin{center}
\textit{Contributed to}  \textit{3rd Les Houches Workshop: Physics at TeV Colliders} \\
\textit{Les Houches, France} \\
\textit{May 26 - June 6, 2003} \\
\end{center}

\section{Introduction}

The precision measurement of the Higgs boson couplings to fermions and gauge bosons is one of the most important goals of 
an $\ee$ linear collider.  These measurements will distinguish between different models of electroweak symmetry breaking,
and can be used to extract parameters within a specific model, such as supersymmetry.  
Most linear collider Higgs studies have
been made assuming a center-of-mass energy of 0.35 TeV, where the Higgsstrahlung cross-section is not too far from its
peak value for Higgs boson masses less than 250~GeV .
Higgs branching fraction measurements with errors of $2-10$\% can be achieved at $\sqrt{s}=0.35$~TeV for many Higgs decay modes, and 
the total Higgs width can be measured with an accuracy of $5-13$\% if the $\sqrt{s}=0.35$~TeV data is combined with $WW$ fusion 
production at $\sqrt{s}=0.50$~TeV\cite{Aguilar-Saavedra:2001rg}.
These measurement errors are very good, but is it possible to do better?  

In the CLIC study of physics at a 3~TeV $\ee$ linear collider it was recognized  that rare Higgs decay modes such as $h\ra\mu^+\mu^-$ could be observed using
Higgs bosons produced through $WW$ fusion\cite{Battaglia:2001vf, Battaglia:2002gq}.  This is possible because
the cross-section for Higgs production through $WW$ fusion  rises with center-of-mass energy, while the design luminosity
of a linear collider also rises with energy.
One doesn't have to wait for 
a center-of-mass energy of 3~TeV, however, to take advantage of this situation.   Already at $\sqrt{s}=1$~TeV the cross-section for Higgs boson production through $WW$ fusion
is two to four times larger than the Higgsstrahlung cross-section at $\sqrt{s}=0.35$~TeV, and the linear collider design luminosity is two times larger at $\sqrt{s}=1$~TeV than at $\sqrt{s}=0.35$~TeV\cite{unknown:2003na}.
Table~\ref{tab:higgs-sample_comparison} summarizes the Higgs event rates at $\sqrt{s}=0.35$ and 1~TeV for several Higgs boson masses. 

\begin{table}[b]
\begin{center}
\caption{Number of inclusive Higgs events assuming an initial state electron polarization of -80\% and integrated luminosities 
of 500 (1000) fb$^{-1}$ for $\sqrt{s}=350$ (1000) GeV.
Effects from beamstrahlung and initial state radiation are included assuming the NLC machine design.  
}
\begin{tabular}{rr|rrrr}
     &  &                  \multicolumn{4}{c}{Higgs Mass (GeV)} \\
 $\sqrt{s}$ (GeV) & $e^+_{\rm pol}$ (\%)  & 120 & 140 & 160 & 200 \\
\hline
& & & & & \\
 350  & 0 &  110280  & 89150    & 69975   &  37385   \\
 350  & +50 &  159115  &  128520   & 100800   &  53775   \\
 1000 & 0 &  386550  &  350690  &  317530 &  259190  \\
 1000 & +50 &  569750  &  516830   & 467900   & 382070   \\

\hline
\end{tabular}
\label{tab:higgs-sample_comparison}
\end{center}
\end{table}

In this report methods for extracting Higgs boson signals at 
a 1~TeV center-of-mass energy $\ee$ linear collider are presented, along with estimates 
of the accuracy with which the Higgs boson cross-section times branching fractions, $\sigma\cdot B_{xx}$, can be measured.  
All results and figures at $\sqrt{s}=1$~TeV assume 1000~fb$^{-1}$ luminosity, -80\% electron polarization, and +50\% positron polarization.

\section{Event Simulation}

The Standard Model backgrounds from all 0,2,4,6-fermion processes and the top quark-dominated 8-fermion processes are generated 
at the parton level using the WHIZARD Monte Carlo\cite{Kilian:2002cg}. 
  In the case of processes such as 
$\ee\ra \ee f \bar{f}$ the photon flux from real beamstrahlung photons is included along with the photon flux from Weisz\"{a}cker-Williams low-$q^2$ virtual photons.
The production of the Higgs boson and its subsequent decay to $b\bar{b}$ and $\tau^+\tau^-$ is automatically included in WHIZARD in the generation of the 4-fermion
processes
$\ee\ra f\bar{f} b\bar{b}$ and $\ee\ra f\bar{f} \tau^+\tau^-$.  For other Higgs decay modes the WHIZARD Monte Carlo is used to simulate $\ee\ra f\bar{f} h$ and
the decay of the Higgs boson is then simulated using PYTHIA\cite{Sjostrand:2000wi}.
The PYTHIA program is also used for final state QED and QCD radiation and for hadronization.
The CIRCE parameterization\cite{Ohl:1997fi} of the
NLC design\cite{unknown:2003na} at $\sqrt{s}=1$~TeV is used to simulate the effects of beamstrahlung.   For
the detector Monte Carlo the SIMDET V4.0 simulation\cite{Pohl:2002vk} of the TESLA detector\cite{Behnke:2001qq} is utilized.

\section{Measurement of $\bf \sigma\cdot B_{xx}$ at $\bf \sqrt{s}=1$~TeV}

Results will be presented for the Higgs decay modes $h\ra b\bar{b},\  WW,\ gg,\ \gamma\gamma$.  The $h\ra c\bar{c}$ decay is not studied 
since a detailed charm-tagging analysis is beyond the scope of this paper;  however it might be interesting for charm-tagging experts 
to pursue this decay mode at $\sqrt{s}=1$~TeV.  The  $h\ra \tau^+\tau^-$ decay is not considered since the neutrinos from the decays of the taus
severely degrade the Higgs mass reconstruction.

Higgs events are preselected by requiring that there be no isolated electron or muon, and that the angle of the thrust axis $\theta_{\rm thrust}$,
visible energy $E({\rm visible})$, and total visible transverse momentum $p_T({\rm visible})$ satisfy
\begin{eqnarray}
	|\cos{\theta_{\rm thrust}}|<0.95,\quad && \nonumber \\
	 100 <E({\rm visible})<400\ {\rm GeV}, && 20<p_T({\rm visible})<500{\rm GeV}.
\end{eqnarray}
Other event variables which will be used in the Higgs event selection include the total visible mass $M({\rm visible})$, the number of charged tracks  $N({\rm chg})$,
the number of large impact parameter charged tracks $N({\rm imp})$, and
the number of jets $N({\rm jet})$ as determined by the PYCLUS algorithm of PYTHIA with parameters MSTU(46)=1 and PARU(44)=5.

\subsection{$\bf h\ra b\bar{b}$ }
\begin{figure}
\begin{center}
\includegraphics[width=7.5cm]{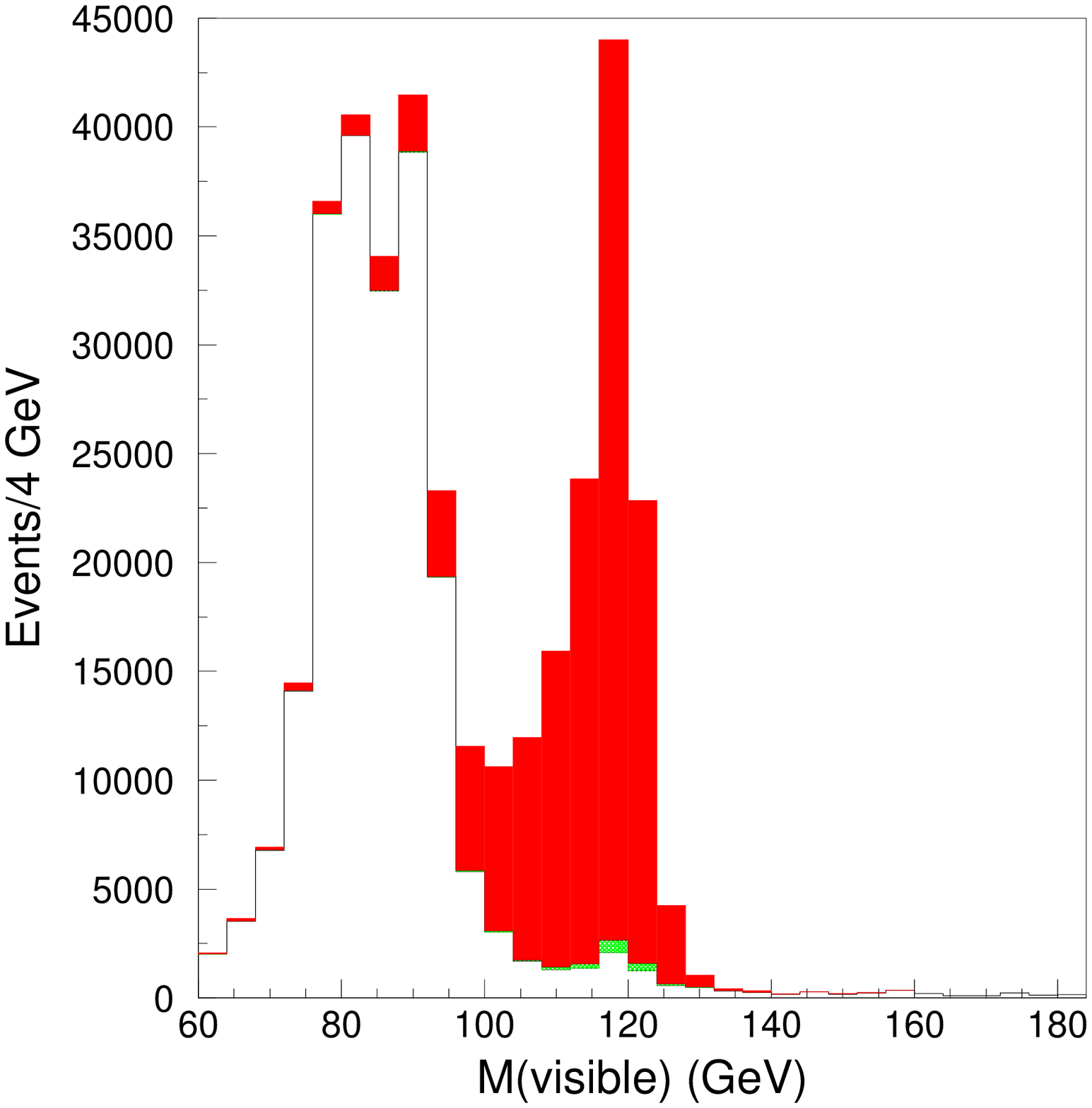} 
\includegraphics[width=7.5cm]{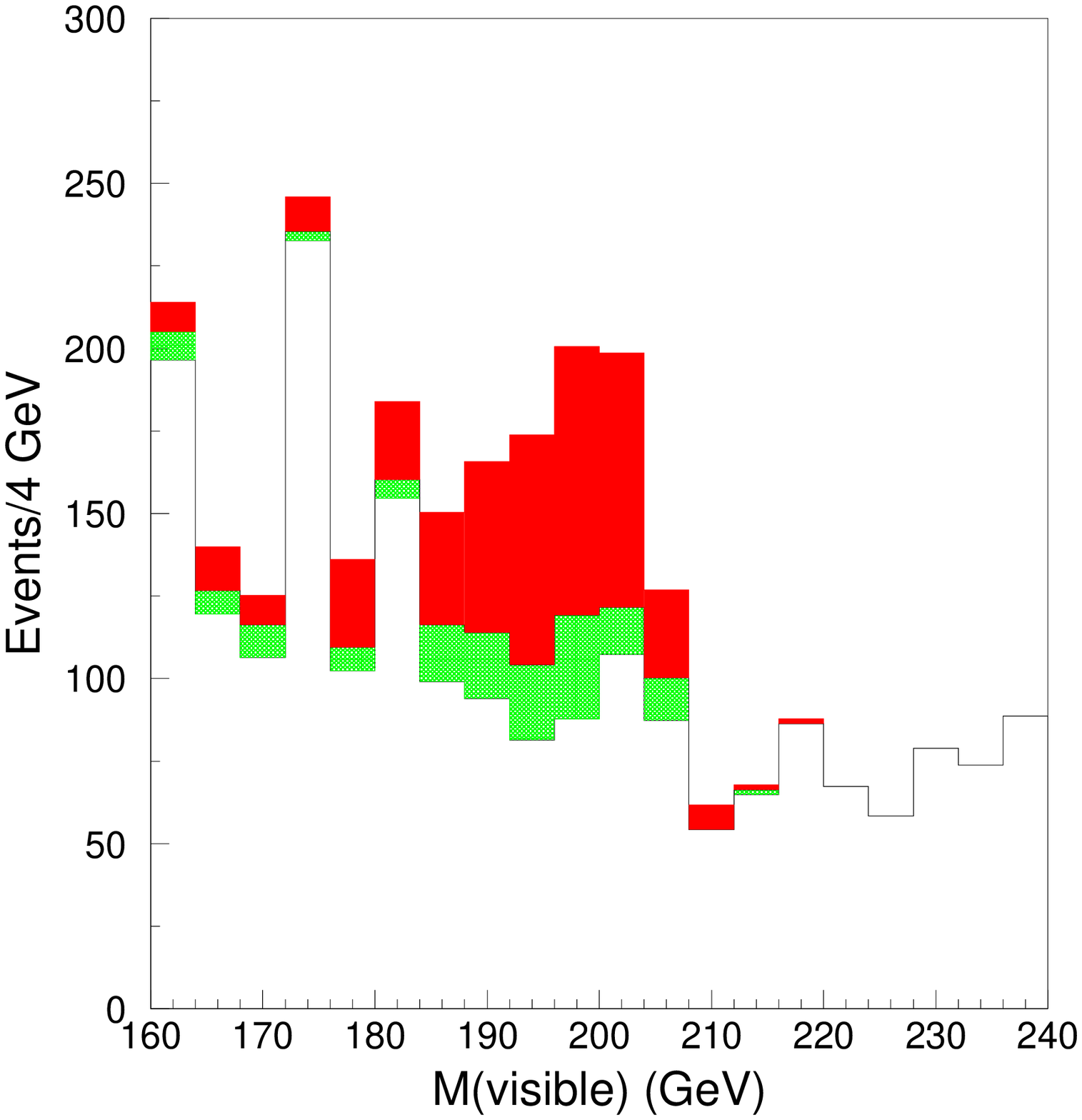} 
\caption{Histograms of $M({\rm visible})$ following $b\bar{b}$ selection cuts for background and signal assuming 
$M_h=120$~GeV (left) and $M_h=200$~GeV (right).  The histograms contain
non-Higgs SM background (white), $h\ra b\bar{b}$ (red) and other Higgs decays (green). 
}
\label{fig-mvis_bb}
\end{center}
\end{figure}

      Decays of  Higgs bosons to $b$ quarks are selected by requiring:
\begin{eqnarray}
      6\le N({\rm chg}) \le 19,\quad &&  7\le N({\rm imp}) \le 19, \nonumber \\
      2\le N({\rm jet}) \le 3,\quad  &&M_h-10\ {\rm GeV}<M({\rm visible})<M_h+6\ {\rm GeV},
\end{eqnarray}
where $M_h$ is the Higgs boson mass measured at $\sqrt{s}=350$~GeV.
Histograms of $M({\rm visible})$ are shown in Fig.~\ref{fig-mvis_bb} assuming Higgs boson masses of 120 and 200~GeV.   
Most of the non-Higgs SM background in the left-hand plot is due to $\ee\ra e\nu W,\ eeZ,\ \nu \nu Z$, while the non-Higgs background in the right-hand plot
is mostly $\gamma\gamma \ra WW$.
The statistical accuracy for cross-section times branching
ratio, $\sigma\cdot B_{bb}$,  is shown in the first row of Table~\ref{tab:higgs-sigma_br}, along with results for $M_h=115,\ 140,\ {\rm and}\ 160$~GeV.    

The Higgs background makes up 1.2\% of the events in the left-hand plot that pass all cuts, and of these
70\% are $c\bar{c}$, 20\%  are $gg$, 5\% are $WW^*$, and 5\% are $ZZ^*$.   The Higgs background is small enough that Higgs branching fraction measurements from
$\sqrt{s}=350$~GeV can be used to account for this background without introducing a significant systematic error.  The non-Higgs background should be calculated with an accuracy
of 1 to 2\% to keep the non-Higgs background systematic error below the statistical error.

\subsection{$\bf h\ra \gamma\gamma$ }

\begin{figure}
\begin{center}
\includegraphics[width=7.5cm]{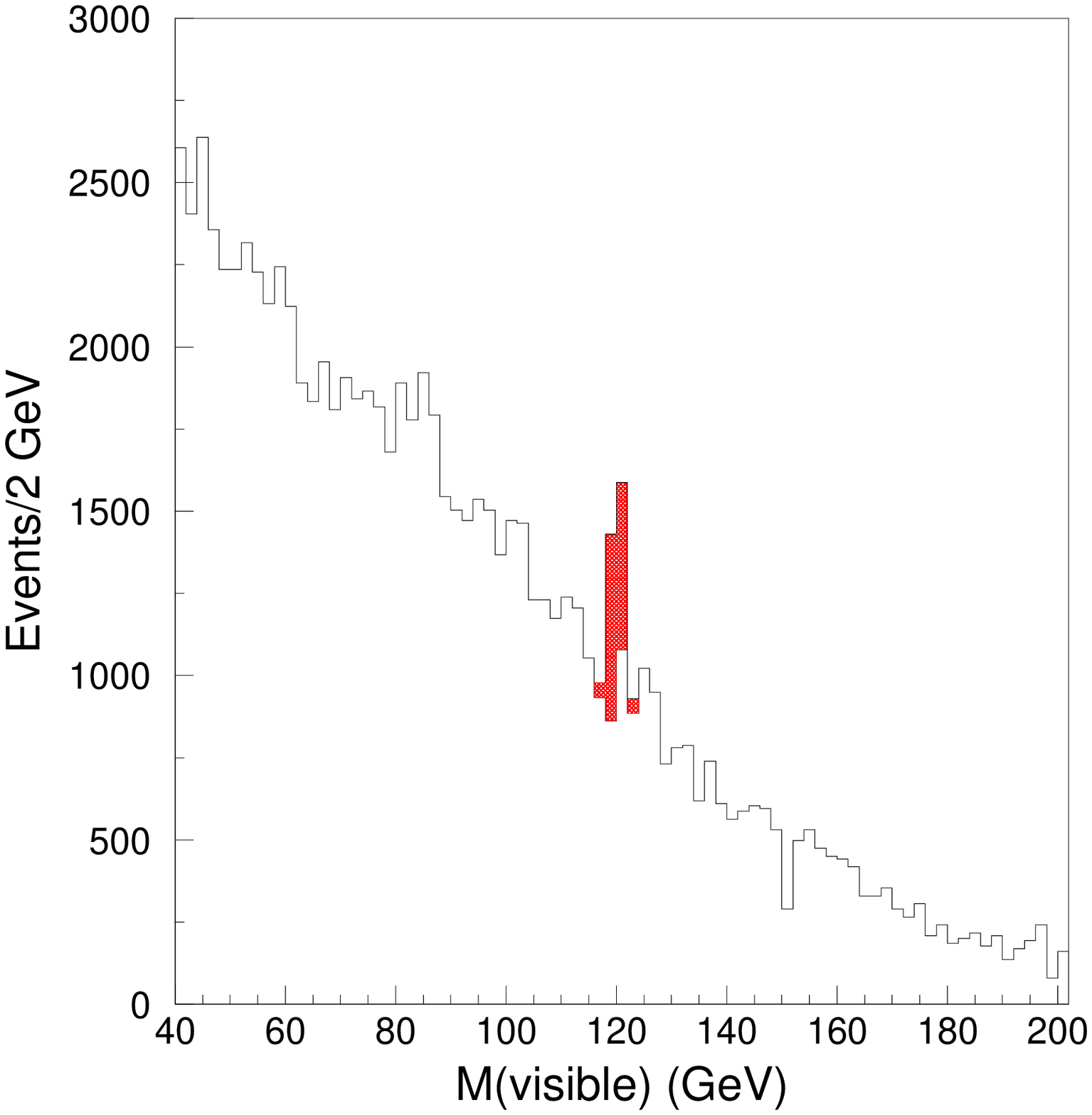} 
\includegraphics[width=7.5cm]{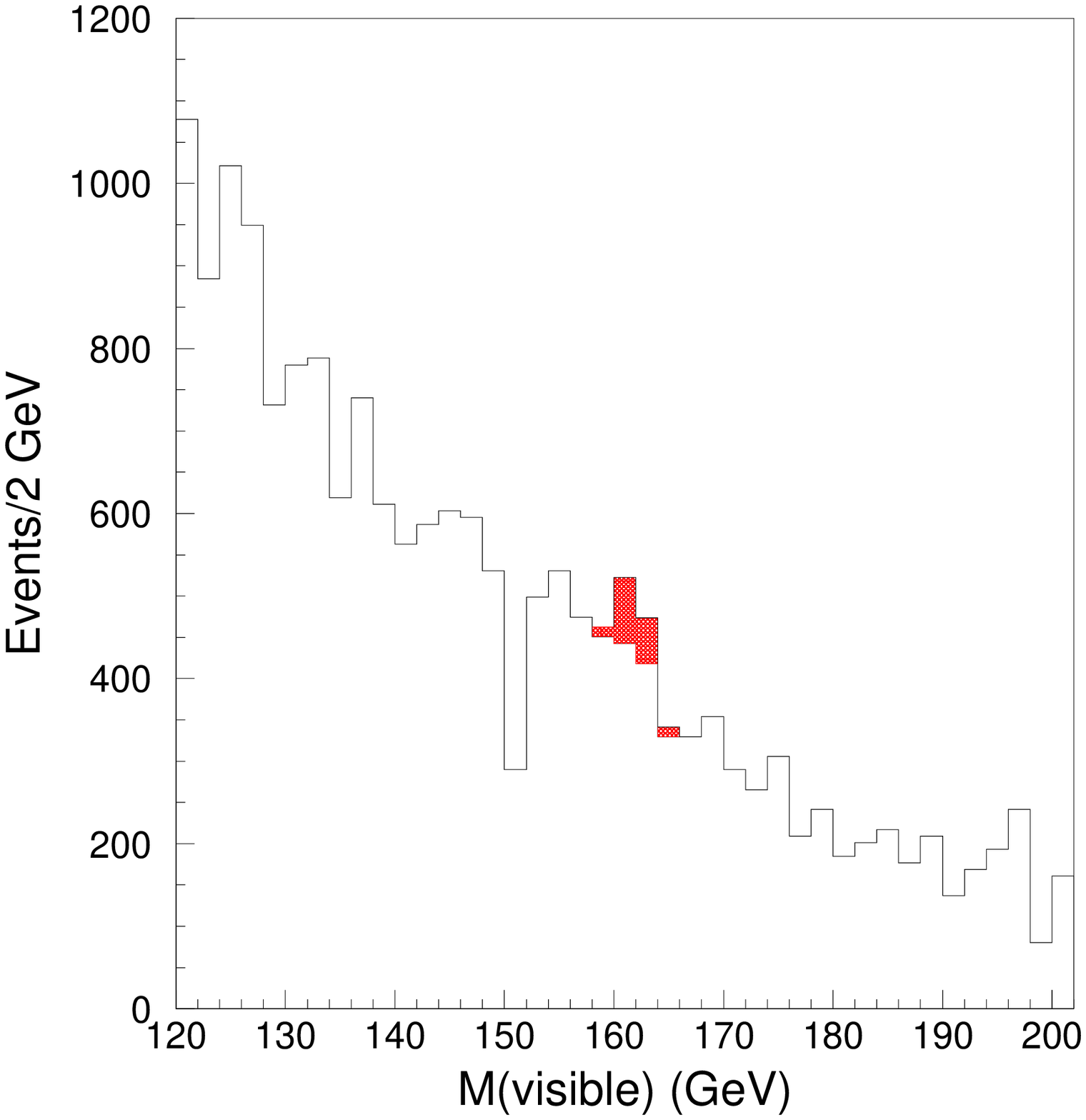} 
\caption{Histograms of $M({\rm visible})$ following $\gamma\gamma$ selection cuts for background and signal assuming 
$M_h=120$~GeV (left) and $M_h=160$~GeV (right).  The histograms contain
non-Higgs SM background (white) and $h\ra \gamma\gamma$ (red). 
}
\label{fig-mvis_aa}
\end{center}
\end{figure}

      Decays of  Higgs bosons to photon pairs are selected by requiring:
\begin{eqnarray}
      N({\rm chg})=0,\quad &&  N({\rm imp})=0, \nonumber \\
      N({\rm jet})=2,\quad  &&M_h-2\ {\rm GeV}<M({\rm visible})<M_h+1\ {\rm GeV}.
\end{eqnarray}
Histograms of $M({\rm visible})$ are shown in Fig.~\ref{fig-mvis_aa} assuming Higgs boson masses of 120 and 160~GeV.   
The SM background is almost entirely $\ee\ra \nu\nu\gamma\gamma$.

\subsection{$\bf h\ra WW,\ \ gg$}

\begin{figure}
\begin{center}
\includegraphics[width=7.5cm]{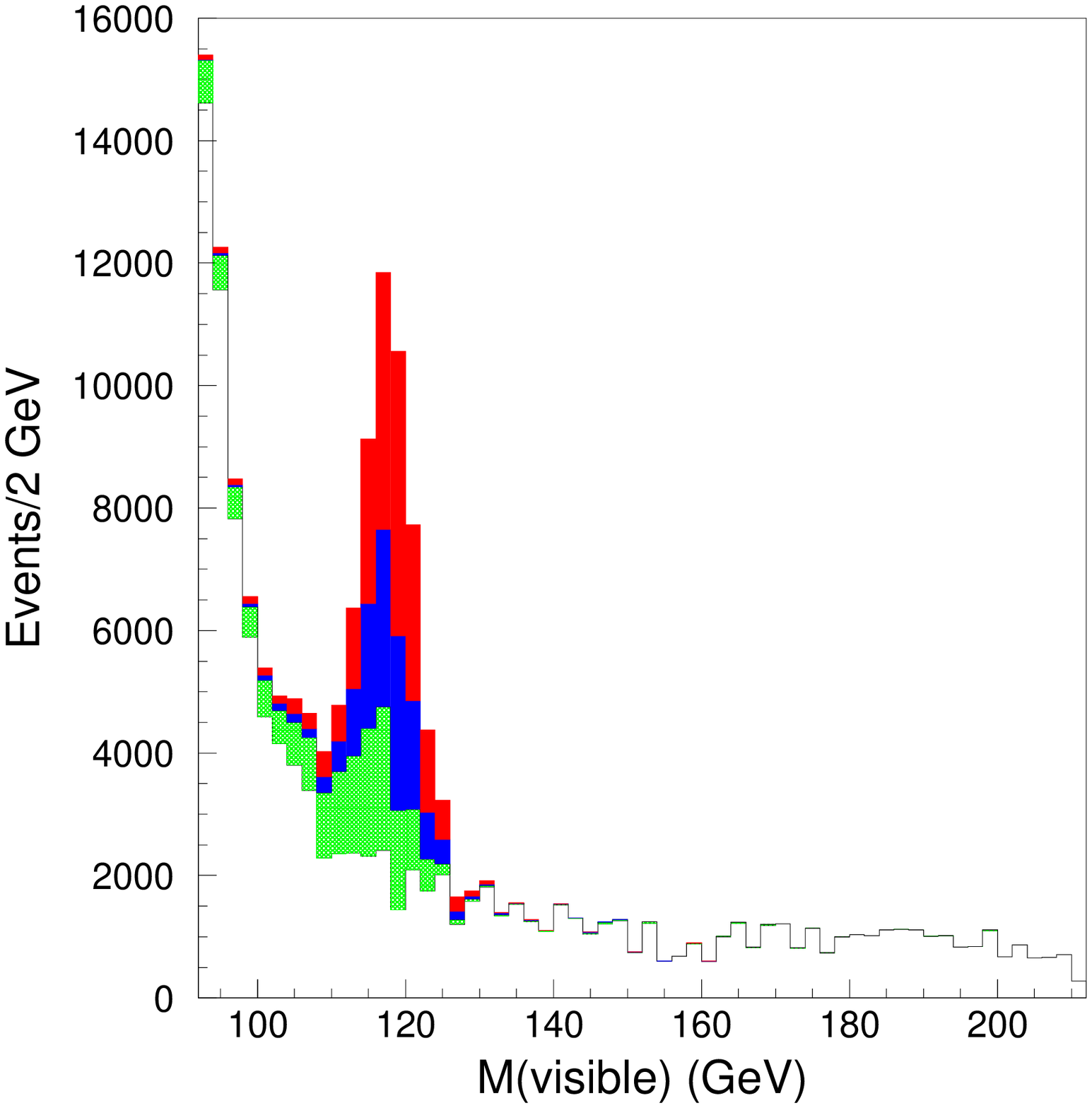} 
\includegraphics[width=7.5cm]{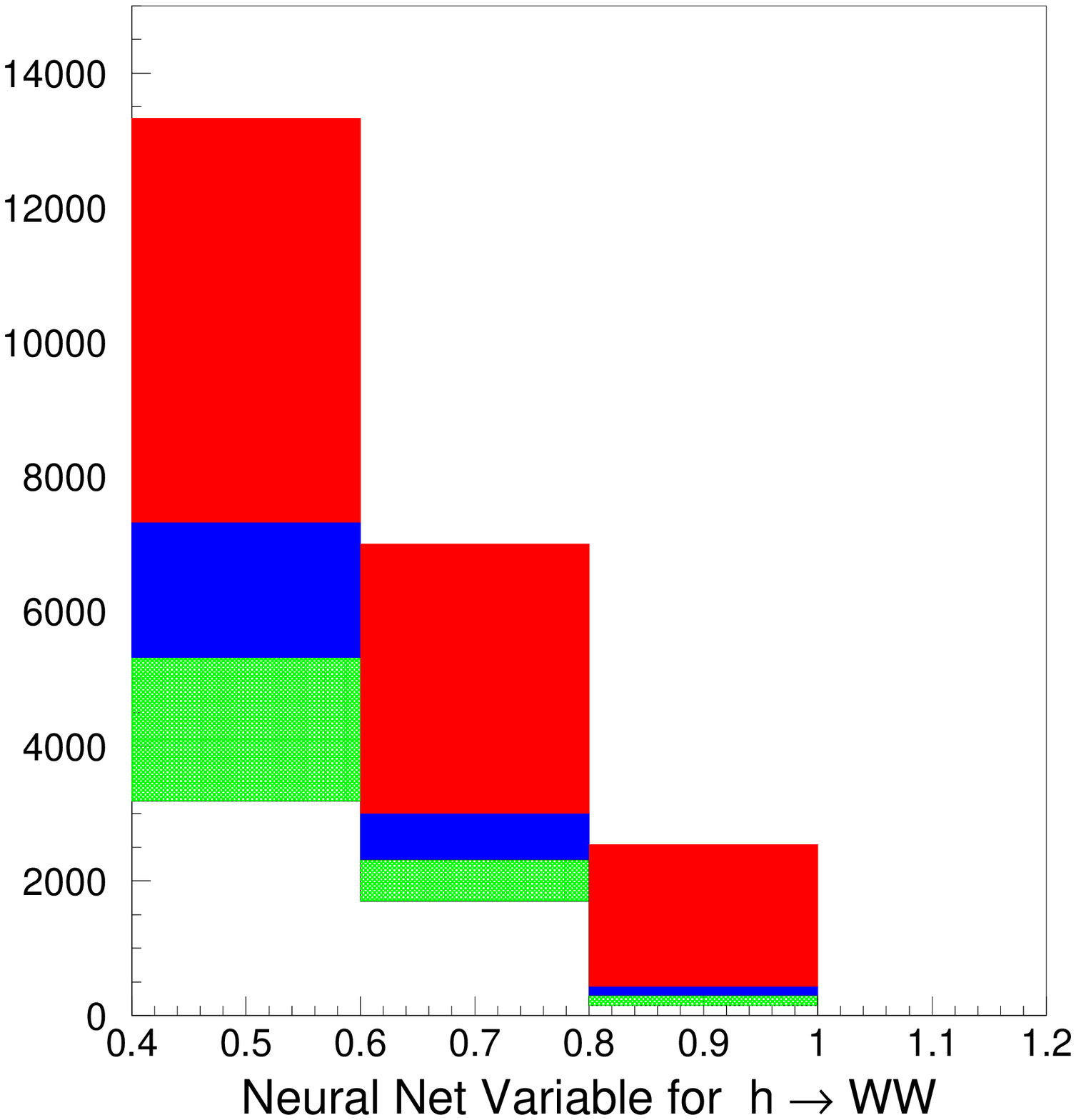} 
\caption{Histograms of $M({\rm visible})$ (left) and the $h\ra WW$ neural net variable (right) following $WW$ selection cuts assuming 
$M_h=120$~GeV.  The histograms contain
non-Higgs SM background (white), $h\ra W\bar{W}$ (red), $h\ra gg$ (blue), and $h\ra b\bar{b},\ c\bar{c},\ ZZ^*$ (green). 
}
\label{fig-mvis_ww}
\end{center}
\end{figure}

      Decays of  Higgs bosons to $WW$ or $WW^*$ are selected by requiring:
\begin{eqnarray}
      16\le N({\rm chg}) \le 44,\quad &&  N({\rm imp}) \le 6, \nonumber \\
      4\le N({\rm jet}) \le 5,\quad  &&M_h-10\ {\rm GeV}<M({\rm visible})<M_h+6\ {\rm GeV}.
\end{eqnarray}
The histogram of $M({\rm visible})$ following the $WW$ cuts is shown in the left-hand side of Fig.~\ref{fig-mvis_ww} for a Higgs boson mass of 120~GeV.   
The non-Higgs SM background is mostly $\ee\ra e\nu W$.  There is also a substantial Higgs boson background  consisting of $h\ra gg$ (63\%), $h\ra bb$ (14\%),
$h\ra cc$ (12\%) and $h\ra ZZ^*$(12\%).
In order to isolate the $h\ra WW$ signal from the other Higgs decay modes, events are forced into 4 jets and a neural net analysis is performed using 
the 4-momentum dot products between pairs of jets and the event variables $E({\rm visible})$, $p_T({\rm visible})$,
$N({\rm chg})$,  $N({\rm imp})$, and $N({\rm jet})$.  The results of this neural net analysis are
shown in the right-hand side of Fig.~\ref{fig-mvis_ww}.

The background from $h\ra b\bar{b},\ c\bar{c},\ ZZ^*$ is small enough that Higgs branching fraction results from $\sqrt{s}=350$~GeV can be used to account for these decays
without introducing significant systematic errors.
However, the contribution from $h\ra gg$ can only be dealt with by measuring
 $\sigma\cdot B_{WW}$ and $\sigma\cdot B_{gg}$ simultaneously.   To that end the decay  $h\ra gg$ is selected by requiring:
\begin{eqnarray}
      11\le N({\rm chg}) \le 49,\quad &&  N({\rm imp}) \le 6, \nonumber \\
      2\le N({\rm jet}) \le 4,\quad  &&M_h-10\ {\rm GeV}<M({\rm visible})<M_h+6\ {\rm GeV}.
\end{eqnarray}
An  $h\ra gg$ neural net analysis is performed with a set of variables identical to that used in the $h\ra WW$ neural net analysis.  The results of the simultaneous fit
of $\sigma\cdot B_{WW}$ and $\sigma\cdot B_{gg}$ for $M_h=115,120,140,160$~GeV are shown in rows 2 and 3 of Table~\ref{tab:higgs-sigma_br}.  For $M_h=200$~GeV
the $h\ra gg$ decay mode is negligible and so
a simultaneous fit of $\sigma\cdot B_{WW}$ and $\sigma\cdot B_{ZZ}$ is made where the $ZZ$ selection cuts are the same as the $WW$ selection cuts and
an $h\ra ZZ$ neural net analysis is performed to separate $h\ra ZZ$ from $h\ra WW$.

\begin{table}[]
\begin{center}
\caption{Statistical accuracies for the measurement of $\sigma\cdot B_{xx}$ for different Higgs decay 
modes $h\ra xx$ 
at $\sqrt{s}=1000$~GeV.
}
\begin{tabular}{l|rrrrr}
     &                    \multicolumn{5}{c}{Higgs Mass (GeV)} \\
   & 115 & 120 & 140 & 160 & 200 \\
\hline
& & & & & \\
 $\Delta (\sigma\cdot B_{bb})/ (\sigma\cdot B_{bb})$ & $\pm 0.003$ &  $\pm 0.004$  & $\pm 0.005$    & $\pm 0.018$  & $\pm 0.090$   \\
 $\Delta (\sigma\cdot B_{WW})/ (\sigma\cdot B_{WW})$ & $\pm 0.021$ &  $\pm 0.013$  & $\pm 0.005$    & $\pm 0.004$  & $\pm 0.005$   \\
 $\Delta (\sigma\cdot B_{gg})/ (\sigma\cdot B_{gg})$ & $\pm 0.014$ &  $\pm 0.015$  & $\pm 0.025$    & $\pm 0.145$  &               \\
 $\Delta (\sigma\cdot B_{\gamma\gamma})/ (\sigma\cdot B_{\gamma\gamma})$ & $\pm 0.053$ &  $\pm 0.051$  & $\pm 0.059$    & $\pm 0.237$  &                \\
$\Delta (\sigma\cdot B_{ZZ})/ (\sigma\cdot B_{ZZ})$ &             &               &                &              & $\pm 0.013$    \\

\hline
\end{tabular}
\label{tab:higgs-sigma_br}
\end{center}
\end{table}

\section{Measurement of Higgs Branching Fractions and the Total Higgs Decay Width}

The measurements of $\sigma\cdot B_{xx}$ in Table~\ref{tab:higgs-sigma_br} can be converted into model independent measurements of Higgs branching fractions and the total Higgs decay width if they
are combined with measurements of the branching fractions $B_{bb}^*$ and $B_{WW}^*$ from $\sqrt{s}=350$~GeV:
\begin{eqnarray}
	B_{xx} &=& (\sigma\cdot B_{xx})(\sigma\cdot B_{WW})^{-1}B_{WW}^*=(\sigma\cdot B_{xx})(\sigma\cdot B_{bb})^{-1}B_{bb}^*             \nonumber \\
	  \Gamma_{tot} & \propto & (\sigma\cdot B_{bb})(B_{bb}^*)^{-1}(B_{WW}^*)^{-1}=(\sigma\cdot B_{bb})^2(\sigma\cdot B_{WW})^{-1}(B_{bb}^*)^{-2}.
\label{eqn-sig_br_to_br}
\end{eqnarray}
The assumed values for the errors on $B_{bb}^*$ and $B_{WW}^*$  are shown in Table~\ref{tab:higgs-assumed_low_energy}.  The errors are taken from the TESLA TDR\cite{Aguilar-Saavedra:2001rg}
when the branching fractions are small.  For large branching fractions, however, it is better to use the direct method\cite{Brient:2002qz} for measuring branching fractions because binomial statistics 
reduce the error by a factor of $\sqrt{1-B_{xx}}$.

Utilizing the relations in Eq.(6) a least squares fit is performed to obtain measurement errors for $B_{bb}$, $B_{WW}$, $B_{gg}$, $B_{\gamma\gamma}$,
and $\Gamma_{tot}$ at a fixed value of $M_h$.   The results are summarized in Table~\ref{tab:higgs-final_errors}.  Compared to branching fraction measurements at 
 $\sqrt{s}=350$~GeV\cite{Aguilar-Saavedra:2001rg} the results of Table~\ref{tab:higgs-final_errors} provide a significant improvement for Higgs decay modes with small branching fractions,
such as $B_{bb}$ for $160<M_h<200$~GeV,  $B_{WW}$ for $115<M_h<140$~GeV and $B_{gg}$ and  $B_{\gamma\gamma}$ for all Higgs masses.

\begin{table}[]
\begin{center}
\caption{Assumed branching fraction errors  for Higgs boson decays to $\ bb\ $ and $WW$  from measuements made at
$\sqrt{s}=350$~GeV  with 500 fb$^{-1}$ luminosity.  
}
\begin{tabular}{l|rrrrr}
     &                    \multicolumn{5}{c}{Higgs Mass (GeV)} \\
   & 115 & 120 & 140 & 160 & 200 \\
\hline
& & & & & \\
 $\Delta B_{bb}^*/B_{bb}^*$ & $\pm 0.015$ &  $\pm 0.017$  & $\pm 0.026$    & $\pm 0.065$  & $\pm 0.340$   \\
 $\Delta B_{WW}^*/B_{WW}^*$ & $\pm 0.061$ &  $\pm 0.051$  & $\pm 0.025$    & $\pm 0.010$  & $\pm 0.025$   \\

\hline
\end{tabular}
\label{tab:higgs-assumed_low_energy}
\end{center}
\end{table}

\begin{table}[]
\begin{center}
\caption{Relative accuracies for the measurement of Higgs branching fractions and the Higgs boson total decay width 
obtained by combining results from Tables \ref{tab:higgs-sigma_br} and \ref{tab:higgs-assumed_low_energy}.
}
\begin{tabular}{l|rrrrr}
     &                    \multicolumn{5}{c}{Higgs Mass (GeV)} \\
   & 115 & 120 & 140 & 160 & 200 \\
\hline
& & & & & \\
 $\Delta B_{bb}/B_{bb}$ & $\pm 0.015$ &  $\pm 0.016$  & $\pm 0.018$    & $\pm 0.020$  & $\pm 0.090$   \\
 $\Delta B_{WW}/B_{WW}$ & $\pm 0.024$ &  $\pm 0.020$  & $\pm 0.018$    & $\pm 0.010$  & $\pm 0.025$   \\
 $\Delta B_{gg}/B_{gg}$ & $\pm 0.021$ &  $\pm 0.023$  & $\pm 0.035$    & $\pm 0.146$  &               \\
 $\Delta B_{\gamma\gamma}/B_{\gamma\gamma}$ & $\pm 0.055$ &  $\pm 0.054$  & $\pm 0.062$    & $\pm 0.237$  &               \\
 $\Delta \Gamma_{tot}/\Gamma_{tot}$ & $\pm 0.035$ &  $\pm 0.034$  & $\pm 0.036$    & $\pm 0.020$  & $\pm 0.050$   \\

\hline
\end{tabular}
\label{tab:higgs-final_errors}
\end{center}
\end{table}

\section{Conclusion}

The couplings of Higgs bosons in the mass range $115<M_h<200$~GeV can continue to be measured as the
energy of an $\ee$ linear collider is upgraded to $\sqrt{s}=1000$~GeV.  The Higgs event rate is so large that
some of the rarer decay modes that were inaccessible at $\sqrt{s}=350$~GeV can be probed at $\sqrt{s}=1000$~GeV, 
such as $h\ra b\bar{b}$ for  $M_h=200$~GeV, and  $h\ra gg,\ \gamma\gamma$  for  $M_h=140$~GeV.  The Higgs physics results
from $\sqrt{s}=1000$~GeV will help provide a more complete picture of the Higgs boson profile.

\vskip 0.7cm
\noindent

\section*{Acknowledgements}

I would like to thank the Les Houches conference organizers for their warm hospitality, and 
I would like to thank John Jaros and
Oliver Buchm\"{u}ller for useful conversations.

\bibliography{biblifile}
\end{document}